# Label-Free Nanoscopy with Contact Microlenses: Super-Resolution Mechanisms and Limitations

**Vasily N. Astratov[1,2*], Farzaneh Abolmaali[1], Aaron Brettin[1,2], Kenneth W. Allen[1], Alexey V. Maslov[3], Nicholaos I. Limberopoulos[2], Dennis E. Walker Jr.[2], Augustine M. Urbas[4]**

[1]*Department of Physics and Optical Science, Center for Optoelectronics and Optical Communication, University of North Carolina at Charlotte, Charlotte, North Carolina 28223-0001, USA*

[2]*Air Force Research Laboratory, Sensors Directorate, Wright Patterson AFB, Dayton, OH 45433, USA*

[3]*Department of Radiophysics, University of Nizhny Novgorod, Nizhny Novgorod, Russia*

[4]*Air Force Research Laboratory, Materials and Manufacturing Directorate, Wright Patterson AFB, Dayton, OH 45433, USA*

*Tel: 1 (704) 687 8131, Fax: 1 (704) 687 8197, E-mail: astratov@uncc.edu

**ABSTRACT**
Despite all the success with developing super-resolution imaging techniques, the Abbe limit poses a severe fundamental restriction on the resolution of far-field imaging systems based on diffraction of light. Imaging with contact microlenses, such as microspheres or microfibers, can increase the resolution by a factor of two beyond the Abbe limit. The theoretical mechanisms of these methods are debated in the literature. In this work, we focus on the recently expressed idea that optical coupling between closely spaced nanoscale objects can lead to the formation of the modes that drastically impact the imaging properties. These coupling effects emerge in nanoplasmonic or nanocavity clusters, photonic molecules, or various arrays under resonant excitation conditions. The coherent nature of imaging processes is key to understanding their physical mechanisms. We used a cluster of point dipoles, as a simple model system, to study and compare the consequences of coherent and incoherent imaging. Using finite difference time domain modeling, we show that the coherent images are full of artefacts. The out-of-phase oscillations produce zero-intensity points that can be observed with practically unlimited resolution (determined by the noise). We showed that depending on the phase distribution, the nanoplasmonic cluster can appear with the arbitrary shape, and such images were obtained experimentally.

## 1. INTRODUCTION

In recent years, a new field of microscopy - super-resolved fluorescence (FL) microscopy - was created and marked by the 2014 Nobel Prize in Chemistry that was awarded to Eric Betzig, Stefan Hell, and William E. Moerner for their pioneering work [1]. Staining biological samples with dyes allows 'highlighting' and making visible some sub-structures; however, it is not always a desirable option. On the other hand, the label-free microscopy methods usually produce images with less brightness and optical contrast, especially in the case of small-scale structures [2]. The far-field resolution of both methods, super-resolved FL and label-free microscopies, is fundamentally limited by the diffraction of light. The diffraction-limited resolution is given by the Abbe's formula, $\lambda/(2NA)$, where $\lambda$ is the illumination wavelength and NA is the numerical aperture of the microscope objective. The resolution can be increased due to a solid immersion lens principle in which the best diffraction-limited resolution can approach $\lambda/(2n_o)$, where $n_o$ is the index of the object space. However, increasing the resolution further requires different approaches such as the use of optical nonlinearity (similar to STED microscopy [1]), optical near-fields (similar to SNOM), far-field superlenses and hyperlenses, or structured illumination in combination with various image processing methods.

In this context, imaging by contact microlenses such as microspheres and microfibers emerged as a simple alternative technique that provides resolution of nanoplasmonic structures on the scale of $\lambda/7$ [3-12]. Attempts to understand the underlying physical mechanisms responsible for super-resolution imaging in this method were made [13]. They were based on unusually sharp focusing properties of small spheres ("photonic nanojets" [14-16]) and on using their resonant properties. However, only a slight increase of the resolution beyond the Abbe formula was observed. Thus, the experimental resolution values ~$\lambda/7$ remains largely unexplained.

In our recent work [17-19], we proposed that the super-resolution can be facilitated by the excitation of the coupled modes in closely spaced nanoscale objects. Examples of such objects include nanoplasmonic arrays [11,12], coupled nanospheres [20], and coupled-cavity arrays [21,22]. If such modes are resonantly excited, they can be imaged in the far-field due to their coupling into the contact microlenses. Imaging with participation of such modes is less studied compared to incoherent imaging [23].

In this work, we study such coherent imaging of coupled modes using a simplified model of point sources with different spatial configurations and phase distributions to represent qualitatively the properties of such modes. Using finite-difference time-domain (FDTD) modeling, we compared coherent and incoherent images and showed that in the former case there are interesting artifacts. It is shown that some features in the coherent images can be observed with the resolution far beyond the classical Abbe limit.

## 2. EXPERIMENTAL SETUP AND GEOMETRY

In Fig. 1 we illustrated the setup and geometry of the experiments based on using microspheres in contact with nanoplasmonic structures. In this example, we used high-index ($n_{sph}$~1.9-2.2) BaTiO$_3$ glass microspheres with diameter $D_{sph}$=8 μm, which were immersed in isopropyl alcohol (IPA) with a refractive index of $n_o$~1.37. Typical nanoplasmonic array contained Au or Al dimers which were not resolvable by conventional microscope with the best objectives. However, the virtual imaging through the microsphere allows one to resolve the minimum at the center of each dimer. The virtual image was obtained in a confocal mode with the Olympus LEXT-OLS4000 microscope operating with 20×(NA=0.6) objective.

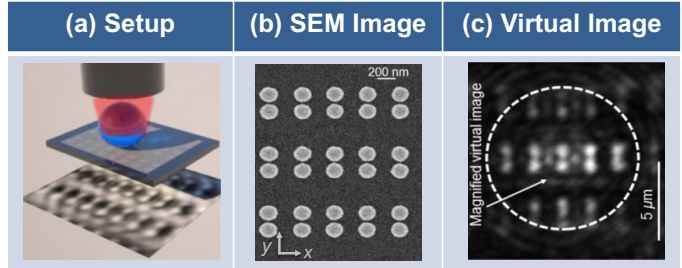

*Figure 1. (a) Set-up, (b) SEM image of the dimer array, and (c) image obtained through a microsphere.*

## 3. NUMERICAL MODELING

The numerical modeling was performed using Lumerical FDTD Solution software. As point sources, we used femtosecond electric dipoles oscillating perpendicular to the object plane with the emission maximum at $\lambda$=405 nm in air. We found that the calculations with microspheres are much more time consuming compared to electromagnetic modeling performed without microspheres, which can be related to multiple reflections of light inside the spheres. To simplify the calculations and represent effects of coherent imaging in a uniform environment, we performed modeling without microspheres. We assumed that the half-space above the object plane has index $n_o$=1.5 to take into account the effective index introduced by the spheres due to solid immersion effect. The images were acquired by an electric field magnitude monitor in the frequency domain and reconstructed using far-field imaging program (developed by Lumerical), which essentially mimics the function of the microscope objective with the maximal effective numerical aperture corresponding in this case to 1.5. The electric field amplitude map is calculated at the virtual image plane identified by the far-field imaging program due to maximal amplitude of the electric field. The mesh size in the medium was $\lambda/(18n_o)$.

As illustrated in Fig. 2(a), we simulated the object illustrated in Fig. 1(b) using 8 point dipoles assembled as 4 pairs (dimers) with the separation ($s$) above the diffraction limit (top row, $s$=200 nm>$\lambda/(2n_o)$=135 nm) and below the diffraction limit (lower row, $s$=100 nm<$\lambda/(2n_o)$=135 nm). The incoherent images of such objects were calculated using a random phase shift between the point sources. Such images presented in Fig. 2(d) illustrate behavior predicted by the Abbe's formula. The image in the top row for $s$=200 nm is resolved, whereas the image in the lower row for $s$=100 nm is not resolved. The appearance of the incoherent images is reminiscent of the experimental image in Fig. 1(c). Even a slightly pronounced peak between the dimers is reproduced by the calculations. This type of peak can be interpreted as an artifact of incoherent imaging.

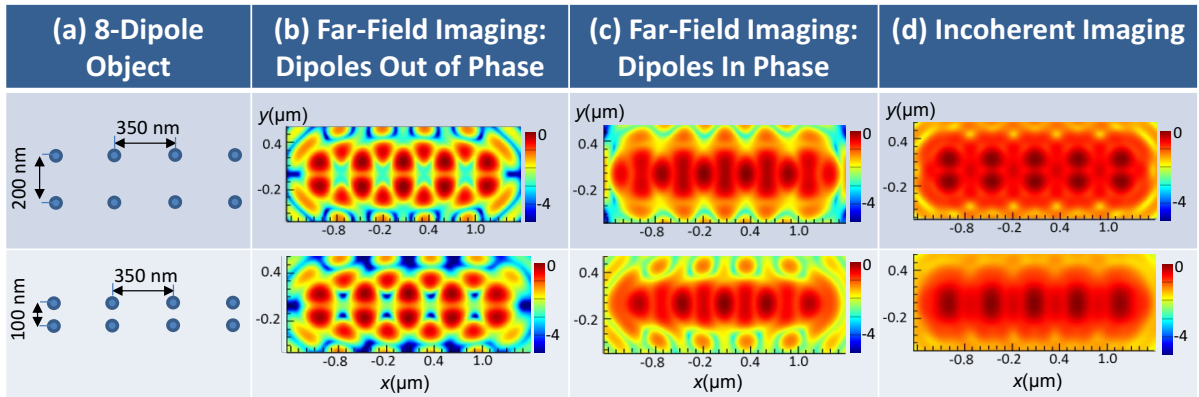

*Figure 2. (a) 8-dipole object, calculated far-field images for point sources in dimers oscillating (b) out-of-phase, (c) in phase, and (d) incoherently.*

The most interesting results were obtained in the case of coherent imaging performed for the out-of-phase and in-phase oscillations of dipoles in dimers, as illustrated in Figs. 2(b) and 2(c), respectively. We found the same tendency which we previously established in the case of 2D modeling [16]. In the out-of-phase case, the images look resolved for any separation (*s*) in dimers. This property can be understood as an artifact of coherent imaging determined by the destructive interference. In principle, it can be viewed as an "infinite" resolution of the zero-intensity point in the middle of each dimer. On the other hand, in the in-phase case, the resolution of dimers is actually worse compared to the incoherent case. This case is accompanied by the appearance of strong peaks between the dimers which are also artifacts of coherent imaging appearing due to constructive interference of the emission of neighboring dimers.

It seems that in most of the cases the imaging of nanoplasmonic structures corresponds to incoherent case. It should be noted, however, that experimentally we do not have an independent control of the modes that can be resonantly excited in such structures. In principle, their excitation may become possible if a coherent laser source with narrow emission line turned out to be in resonance with the coupled modes of the plasmonic or photonic clusters. An example of a situation which strongly suggests such a possibility is illustrated in Fig. 3.

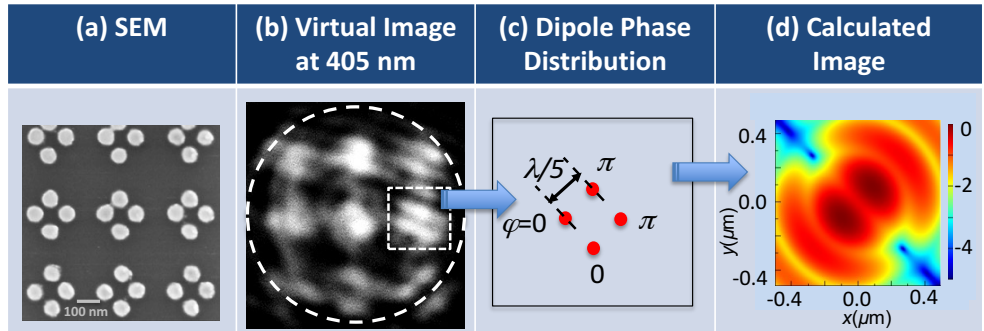

*Figure 3. (a) SEM image of the array of 4-circle objects, (b) virtual image at 405 nm acquired through a 5 μm $SiO_2$ sphere, (c) 4-dipole object with the phase distribution indicated, and (d) calculated far-field image reminiscent of the experimental image in (b).*

The diagonal streaks in Fig. 3(b) can be explained by the excitation of the coupled states of such plasmonic clusters with the characteristic phase distribution illustrated in Fig. 3(c). The calculated image of a single cluster of 4 dipoles with such phase distribution reminds the experimental image of such clusters presented in Fig. 3(b).

**4. CONCLUSIONS**

We performed electromagnetic modeling of several closely-spaced point sources without microspheres. It is shown that the imaging of subwavelength structures is dramatically different in the coherent case compared to the incoherent case. Incoherent imaging correctly displays the shape of the objects, however even this case is not free from artifacts. Coherent imaging is dramatically different, especially for subwavelength structures. If different parts of the nanoscale clusters oscillate out-of-phase, the gap between such parts usually can be observed with "infinite" resolution that does not contradict the classical Abbe limit. The coherent imaging has various artifacts. The role of these artifacts in imaging is dubious. On one hand, they distort the optical images. On the other hand, they can be used for precise localization of some objects and they can also play some part in the experimental quantification of resolution in such structures. Usually, such quantification is developed for incoherent imaging. However, if coherent imaging becomes involved, it needs to be carefully taken into account in order to determine the resolution. This work shows that selective excitation of the coupled states of closely spaced nanoplasmonic and nanophotonic objects can open new ways of their imaging where extraordinary precise positional information can be obtained. In our future work, we will perform similar studies for imaging through microspheres where additional resolution advantages become possible due to the optical magnification and collection of the optical near-fields by the contact microlens.

**ACKNOWLEDGMENTS**
This work was supported by Center for Metamaterials, an NSF I/U CRC, award number 1068050. Also, this work was sponsored by the Air Force Research Laboratory (AFRL/RYD, AFRL/RXC) through the AMMTIAC contract with Alion Science and Technology and the MCF II contract with UES, Inc.

# REFERENCES

[1] S.W. Hell: Nanoscopy with focused light, *Ann. Phys. (Berlin)*, vol. 527, pp. 423-445, Aug. 2015.
[2] S. Weisenburger and V. Sandoghdar: Light microscopy: an ongoing contemporary revolution, *Contemp. Phys.*, vol. 56, pp. 123-143, April 2015.
[3] Z. Wang, W. Guo, L. Li, B. Luk'yanchuk, A. Khan, Z. Liu, Z. Chen, and M. Hong: Optical virtual imaging at 50 nm lateral resolution with a white-light nanoscope, *Nat. Commun.*, vol 2, 218, March 2011.
[4] V.N. Astratov and A. Darafsheh: Methods and systems for super-resolution optical imaging using high-index of refraction microspheres and microcylinders, US Patent US20140355108, 2014 (priority on June 7, 2012); http://www.freepatentsonline.com/20140355108.pdf
[5] A. Darafsheh, G.F. Walsh, L. Dal Negro, and V.N. Astratov: Optical super-resolution by high-index liquid-immersed microspheres, *Appl. Phys. Lett.*, vol. 101, 141128, Oct. 2012.
[6] L. Li, W. Guo, Y, Yan, S. Lee and T. Wang: Label-free super-resolution imaging of adenoviruses by submerged microsphere optical nanoscopy, *Light: Science & Applications*, vol. 2, e104, Sept. 2013.
[7] H. Yang, N. Moullan, J. Auwerx, and M.A.M. Gijs: Super-resolution biological microscopy using virtual imaging by a microsphere nanoscope, *Small*, vol. 10, pp. 1712-1718, May 2014.
[8] A. Darafsheh, N.I. Limberopoulos, J.S. Derov, D.E. Walker, Jr., and V.N. Astratov: Advantages of microsphere-assisted super-resolution imaging technique over solid immersion lens and confocal microscopies, *Appl. Phys. Lett.*, vol. 104, 061117, Feb. 2014.
[9] K.W. Allen, N. Farahi, Y. Li, N.I. Limberopoulos, D.E. Walker Jr., A.M. Urbas, and V.N. Astratov: Super-resolution imaging by arrays of high-index spheres embedded in transparent matrices, IEEE Aerospace and Electronics Conference, in *Proc. NAECON* 2014, Dayton, Ohio, June 2014, pp. 50-52.
[10] K.W. Allen: Waveguide, photodetector, and imaging applications of microspherical photonics, Ph.D. dissertation (University of North Carolina at Charlotte, 2014), Chapter 4: Super-resolution imaging through arrays of high-index spheres embedded in transparent matrices, pp. 98-122, Oct. 2014. http://gradworks.umi.com/36/85/3685782.html
[11] K.W. Allen, N. Farahi, Y. Li, N.I. Limberopoulos, D.E. Walker Jr., A.M. Urbas, V. Liberman, and V.N. Astratov: Super-resolution microscopy by movable thin-films with embedded microspheres: Resolution analysis, *Ann. Phys. (Berlin)*, vol. 527, pp. 513-522, Aug. 2015.
[12] K.W. Allen, N. Farahi, Y. Li, N.I. Limberopoulos, D.E. Walker Jr., A.M. Urbas, V.N. Astratov: Overcoming the diffraction limit of imaging nanoplasmonic arrays by microspheres and microfibers, *Opt. Express*, vol. 23, pp. 24484-24496, Sep. 2015.
[13] T.X. Hoang, Y. Duan, X. Chen, G. Barbastathis: Focusing and imaging in microsphere-based microscopy, *Opt. Express*, vol. 23, pp. 12337-12353, May 2015.
[14] V.N. Astratov, A. Darafsheh, M.D. Kerr, K.W. Allen, N.M. Fried, A.N. Antoszyk, and H.S. Ying: Photonic nanojets for laser surgery, *SPIE Newsroom*, March 2010; http://spie.org/x39280.xml.
[15] A. Darafsheh and V.N. Astratov, Periodically focused modes in chains of dielectric spheres, *Appl. Phys. Lett.*, vol. 100, 061123, Feb. 2012.
[16] K.W. Allen, A. Darafsheh, F. Abolmaali, N. Mojaverian, N.I. Limberopoulos, A. Lupu, and V.N. Astratov: Microsphere-chain waveguides: Focusing and transport properties, *Appl. Phys. Lett.*, vol. 105, 021112, July 2014.
[17] A.V. Maslov and V.N. Astratov: Imaging of sub-wavelength structures radiating coherently near microspheres, *Appl. Phys. Lett.*, vol. 108, 051104, Feb. 2016.
[18] V.N. Astratov, K.W. Allen, N. Farahi, Y. Li, N.I. Limberopoulos, D.E. Walker Jr., A.M. Urbas, V. Liberman and M. Rothschild: Optical nanoscopy with contact microlenses overcomes the diffraction limit, *SPIE Newsroom*, February 11 (2016). http://spie.org/newsroom/6314-optical-nanoscopy-with-contact-microlenses-overcomes-the-diffraction-limit
[19] V.N. Astratov, A.V. Maslov, K.W. Allen, N. Farahi, Y. Li, A. Brettin, N.I. Limberopoulos, D.E. Walker Jr., A.M. Urbas, V. Liberman, and M. Rothschild: Fundamental limits of super-resolution microscopy by dielectric microspheres and microfibers, in *Proc. of SPIE*, vol. 9721, 97210K, 7 pp., March 2016.
[20] P. von Olshausen and A. Rohrbach: Coherent total internal reflection dark-field microscopy: label-free imaging beyond the diffraction limit, *Opt. Lett.*, vol. 38, pp. 4066-4069, Oct. 2013.
[21] A.D. Bristow, V.N. Astratov, R. Shimada, I.S. Culshaw, M.S. Skolnick, D.M. Whittaker, A. Tahraoui, and T.F. Krauss: Polarization conversion in the reflectivity properties of photonic crystal waveguides, *IEEE J. of Quantum El.*, vol. 38, pp. 880-884, July 2002.
[22] A.D. Bristow, D.M. Whittaker, V.N. Astratov, M.S. Skolnick, A. Tahraoui, T.F. Krauss, M. Hopkinson, M.P. Croucher, G.A. Gehring: Defect states and commensurability in dual-period $Al_xGa_{1-x}As$ photonic crystal waveguides, *Phys. Rev.* B, vol. 68, 033303, July 2003.
[23] A. Lipson, S.G. Lipson, and H. Lipson, in *Optical Physics,* New York: Cambridge University Press, 4[th] Edition, 2011.